\documentclass[sigconf]{acmart}
\setcopyright{none}

\pdfoutput=1
\acmDOI{} 

\acmISBN{} 

\acmConference[]{}{}{}
\acmYear{}
\copyrightyear{}

\usepackage{epstopdf}
\usepackage{dblfloatfix}
\usepackage{booktabs}
\usepackage{enumitem}
\usepackage{url}
\usepackage[normalem]{ulem}
\usepackage{subcaption}
 \usepackage{algorithmicx}
 \usepackage{algpseudocode}
 \usepackage[ruled, linesnumbered]{algorithm2e}
\usepackage{color}
\usepackage{adjustbox}
\usepackage{caption} 
\usepackage{framed}

\usepackage{multirow}
\usepackage{color, colortbl}

\newcommand{\REAP}{\textit{REAP~}}
\begin{document}
	




\title{\vspace{-3mm}\textit{REAP}: Runtime Energy-Accuracy Optimization for\\ 
Energy 
Harvesting 
IoT Devices }

{\author{{Ganapati Bhat$^1$, Kunal Bagewadi$^2$, Hyung Gyu Lee$^3$ 
and 
Umit Y. Ogras$^1$} \\
	{$^1$School of Electrical Computer and Energy Engineering, 
		ASU, Tempe, AZ\\
		$^1$School of Computing, Informatics, and Decision Systems Engineering, 
		ASU,
		Tempe, AZ\\
		$^2$ Deagu University, Daegu, Korea\\
		Contact Email: gmbhat@asu.edu}
}}

\begin{abstract}

\vspace{-1mm}
The use of wearable and mobile devices for health monitoring and activity 
recognition applications is increasing rapidly. 
These devices need to maximize their accuracy and active time 
under a tight energy budget imposed by battery and small form-factor 
constraints. 
This paper considers energy harvesting devices 
that run on a limited energy budget to recognize user activities over a given 
period. 
We propose a technique to co-optimize the accuracy and active time by 
utilizing multiple design points with different energy-accuracy trade-offs. 
The proposed technique switches between these design points at runtime 
to maximize a generalized objective function under tight harvested energy 
budget constraints. 
We evaluate the proposed approach experimentally using a custom hardware
prototype and fourteen user studies.
The proposed approach achieves both 46\% higher 
expected accuracy and 66\% longer active time
compared to the highest performance design point.


\end{abstract}

\maketitle

\vspace{-3mm}
\section{Introduction}
\vspace{-0.5mm}
Wearable low-power internet-of-things (IoT) devices 
enable a wide range of health monitoring, activity tracking, and wide area sensing 
applications~\cite{liu2017gazelle,lara2013survey,bonomi2009detection}. 
These devices must stay on for as long as possible to observe the user activities. 
At the same time, they have to provide the maximum quality of service, such activity recognition accuracy. 
These two objectives compete with each other since higher accuracy comes at the cost of larger energy consumption. 
Since weight and form-factor constraints prohibit large batteries, feasibility 
of these devices depends critically on optimizing the energy-accuracy trade-off 
optimally at runtime.



\vspace{1mm}
Widely used dynamic power management techniques optimize the power-performance trade-off 
by switching between different power states at runtime~\cite{cochran2011pack}. 
High-performance states are used to execute computationally heavy 
workloads at the expense of larger power consumption. 
In contrast, low-performance states are used during light workloads to save 
power. 
In analogy, energy-accuracy trade-off in self-powered devices can be optimized by utilizing multiple design points. 
\textcolor{black}{This is a challenging proposition 
since characterizing the accuracy analytically 
is much harder than developing power consumption and performance models. 
For example, we consider an activity recognition application, where a wearable device infers the user activities, 
such as jogging, by processing motion sensor data. 
The recognition accuracy is a strong function of the users. 
Hence, energy-accuracy optimization of requires 
 \textit{user studies} and \textit{optimally chosen design points}, 
in addition to a \textit{runtime optimization algorithm} 
that utilizes multiple design points.}

\vspace{1mm}
This paper presents a Runtime Energy-Accuracy oPtimization framework (\textit{REAP}) for energy-constrained IoT devices. 
While our framework is general, we focus on health and activity monitoring applications 
where a  wearable device processes sensor inputs to infer user activities. 
The recognized activities are sent
to a gateway, such as a smartphone, for further processing.
\REAP co-optimizes the accuracy and active time under a tight energy budget.
This optimization is enabled by the following three contributions.

\noindent \textbf{User studies for accuracy evaluation:}  
We perform experiments with 14 users to recognize six activities:
\textit{sit}, \textit{stand}, \textit{walk}, \textit{jump}, \textit{drive}, 
\textit{lie down} and \textit{transitions} among them. 
During these experiments, we collect 3-axis accelerometer and stretch sensor data. 
We obtain a total of \textcolor{black}{3553 activity windows} from these 
experiments. 
After labeling, we utilize this data for evaluating the accuracy 
of the human activity classifiers used in this work.  

\noindent \textbf{Pareto-optimal design points:} 
A common baseline in activity monitoring applications is to obtain 
a classifier with the highest recognition accuracy. 
High accuracy is obtained by using a sophisticated set of sensors, 
features, and classification algorithms, 
all of which imply a larger energy consumption, hence, lower active time. 
Other design points can be obtained by reducing the number of sensors and feature set to save energy. 
In turn, the energy savings lead to longer active time under a given harvested energy budget, albeit with lower accuracy. 
\textit{To enable this work, we implemented 24 design points (DPs) with varying energy-accuracy trade-offs} on our hardware prototype. 
Among them, we choose five Pareto-optimal DPs as our primary designs used at runtime. 
We provide detailed execution time and power consumption breakdown for sensing, feature generation and processing steps 
for each of these five DPs. 

\noindent \textbf{Runtime optimization algorithm:} 
Given an energy budget, two fundamental objectives are to maximize the recognition accuracy 
and the amount of time the device is on, i.e., the active time. 
We first formulate this co-optimization problem assuming that 
there are $N$ design points with different energy-accuracy trade-offs. 
We define a general objective function that enables us to tune the importance of active time versus recognition accuracy. 
Then, we propose an efficient runtime algorithm that determines how much each 
DP should be used so as to optimize the accuracy-active time 
trade-off. 
Our solution reveals the amount of time the device must operate in each of these design points. 

\vspace{1mm}
Experimental results using a custom prototype based on TI Sensortag~\cite{TI_sensortag} IoT board 
show that \textit{REAP} outperforms all static design points under 
a range of energy budget constraints. 
\REAP achieves both 46\% higher expected accuracy and 
66\% longer active time compared to the highest performance DP.
\textit{REAP} also achieves comparable active time to the lowest energy design 
points 
while providing significantly higher expected accuracy. 
This makes \textit{REAP} suitable for use in a wide range of energy harvesting profiles. 



\textit{The major contributions of this paper are as follows:}
\begin{itemize}
\vspace{-2mm}
\item A runtime technique to co-optimize the accuracy and active time 
of energy-harvesting IoT devices.

\item Pareto-optimal design points with varying energy-accuracy 
trade-offs for human activity recognition (HAR). 

\item Experiments on a custom prototype with 14 user studies 
that show significant improvements both in expected accuracy and active time compared to static design points.

\end{itemize}

\vspace{-1mm}
In the rest, Section~\ref{sec:related_work} reviews the related work. 
Section~\ref{sec:problem} and Section~\ref{sec:design_points} present 
the accuracy-active time optimization problem and DPs used in this paper, respectively.
Finally, Section~\ref{sec:experiments} presents the experimental results 
and Section~\ref{sec:conclusions} concludes the paper.

\vspace{-1mm}
\section{Related Work} \label{sec:related_work}
\vspace{-0.5mm}

Energy harvesting for IoT devices has recently received significant attention 
due to their small form-factor and low capacity batteries~\cite{sudevalayam2011energy,jayakumar2014powering}. 
These devices can be broadly categorized into two classes. 
The first class of devices rely solely on harvested energy and turn off when no energy is 
harvested~\cite{shenck2001energy} 
%
The second class of devices uses a small battery as a backup 
to extend the active time~\cite{kansal2007power,vigorito2007adaptive,buchli2014dynamic}. 
These approaches manage the power consumption of the device such that the total 
energy consumed over a finite horizon is equal to the harvested energy. 
This ensures a long device lifetime without battery replacement or manual 
charging.
\textcolor{black}{\textit{REAP} is applicable to all devices that operate under 
a fixed energy budget}.

Intermittent nature of ambient energy sources necessitates the development of 
energy allocation and duty cycling 
algorithms~\cite{buchli2014dynamic,shaikh2016energy,vigorito2007adaptive,bhat2017near}.
For example, the work in~\cite{kansal2007power} uses linear programming to determine the duty 
cycle of the application as a function of the harvested energy. 
Similarly, the algorithm in~\cite{vigorito2007adaptive} uses a linear quadratic 
controller to assign the duty cycle of the device while maintaining a set 
battery level. 
An algorithm for dynamic power and energy management of an energy-harvesting 
node for long-term energy-neutral operation is presented 
in~\cite{buchli2014dynamic}.
However, these approaches choose between on- and off power states, which leads to sub-optimal operation.
%
Furthermore, they do not have a notion of accuracy or any concrete application, 
unlike our work.

Human activity and health monitoring using wearable devices 
have been an active area of research due to their potential benefits to sports, 
patients with movement disorders, and 
elderly~\cite{jafari2007physical,espay2016technology,lara2013survey,bhat2018online}.
A recent work proposes a wearable system for mobile analysis of running using motion sensors~\cite{liu2017gazelle}. 
The authors selectively identify the best sampling points to maintain high accuracy while reducing sensing and analysis energy overheads.
The work in~\cite{jafari2007physical} presents a framework to 
detect falls by using a wearable device equipped with accelerometers. 
Authors in~\cite{bonomi2009detection} design a classifier that detects physical 
activity using a body-worn accelerometer. 
While this study offers an accurate classifier for human activity recognition, it cannot sustain 
operation under tight energy budget constraints. 
Based on this observation, we find Pareto-optimal design points for the HAR application
that offer varying levels of accuracy and energy consumption. Then, we use 
these design points to maximize the expected accuracy of HAR.

\fontdimen2\font=0.72ex
In summary, we present a unique combination of (1) a
runtime energy-accuracy optimization technique, and 
(2) experimental evaluation with five concrete design points for HAR. 
\textit{We will release our detailed power--performance characterizations and 
user subject data to stimulate research in these areas}.


%

%
%
%

\vspace{-2mm}
\section{\hspace{-3mm} \large{Runtime Energy-Accuracy Optimization}}  \label{sec:problem}


\subsection{Preliminaries}
\vspace{-0.5mm}

We consider human activity monitoring applications implemented on energy-constrained IoT devices.
We denote the period over which the total energy budget is provided as $T_P$, as summarized in Table~\ref{tab:problem_symbols}.
\textit{REAP} computes the energy allocations at runtime with a period of $T_P$, which is set to one hour in our experiments.
If the energy consumption over this period exceeds the amount of harvested energy and remaining battery level,
the device powers down and misses user activity.
Hence, our goal is to maximize the active time and the
expected accuracy over a given period $T_P$.

Suppose that the IoT device can operate at $N$ distinct DPs.
The recognition accuracy achieved by design point $i$ is denoted by $a_i$,
while the corresponding power consumption is given as $P_i$ for $1\leq i \leq N$.
In addition to these design points, we denote the time that the device remains off as $t_{off}$.
Finally, the power consumption during the off period,
which is due to the energy harvesting and the battery charging circuitry, is denoted by $P_{off}$.

\renewcommand{\arraystretch}{0.90}
\begin{table}[]
	\caption{\small{Summary of symbols used in the optimization problem.}}
	\label{tab:problem_symbols}
	\vspace{-3mm}
	\begin{tabular}{@{}cc|cc@{}}
		\toprule
		Symbol & Description & Symbol & Description \\ \midrule
		$T_P$ & Activity period & $J(t)$ & Objective function \\ \midrule
		$E_b$ & Energy budget & $t_i$ & Active time of $DP_i$ \\
		\midrule
		$a_i$ & \begin{tabular}[c]{@{}c@{}}Recognition accuracy\\ of
		$DP_i$\end{tabular} & $\alpha$ &
		\begin{tabular}[c]{@{}c@{}}Accuracy-active time\\ trade-off
		parameter\end{tabular} \\ \midrule
		$P_{off}$ & \begin{tabular}[c]{@{}c@{}}Power consumption\\ in the off
		state\end{tabular} & $P_i$ & \begin{tabular}[c]{@{}c@{}}Power
		consumption\\ of $DP_i$\end{tabular} \\ \bottomrule
	\end{tabular}
\vspace{-3mm}
\end{table}
\vspace{-1mm}

\subsection{Optimization Problem Formulation}
In a given activity period, the system may operate at different design points,
resulting in varying levels of active time and accuracy.
Let $t_i$ denote the amount of time DP $i$ is utilized during $T_P$.
The active time of the device is simply given by the sum of the active times of each DP: $\sum_{i=1}^N t_i$.
Likewise, the expected accuracy over the activity period can be expressed as 
$E\{a\} = \frac{1}{T_P}\sum_{i=1}^{N} a_it_i$.
%
%
The expected accuracy is a useful metric that incorporates both active time and accuracy,
but it does not allow emphasis of one over the other.
Therefore, we define a generalized cost function and solve the following optimization problem:
\vspace{-2mm}
%
%
\begin{align}\label{eq:objective}
& \text{$maximize$} & J(t) = \frac{1}{T_P}\sum_{i=1}^{N} a_i^\alpha t_i &   \\
\label{eq:constraint1}
& \text{$subject~to$} &  t_{off} + \sum_{i=1}^{N} t_i &= T_P
\\\label{eq:constraint2}
& \text{$and$} &  P_{off}t_{off} + \sum_{i=1}^{N}P_it_i \hspace{0.5mm} & \leq E_b
\\\label{eq:constraint3}
&& t_i &\geq 0~~~~\hspace{1mm} 0\leq i \leq N
\end{align}
\vspace{-5mm}

\noindent \textbf{Objection function $J(t)$:}
The parameter $\alpha$ in Equation~\ref{eq:objective}
enables a smooth trade-off between the active time and accuracy.
When $\alpha = 1$, the objective function reduces to the expected accuracy.
Similarly, when $\alpha=0$, the objective function reduces to total active time.
In general, the objective function gives a higher weight to the active time when $\alpha < 1$.
In contrast, the design points with higher accuracy are preferred for $\alpha > 1$.

\noindent\textbf{Constraints:}
The constraint given in Equation~\ref{eq:constraint1} states that the
sum of the active times and off period equals to the overall period $T_P$.
Similarly, Equation~\ref{eq:constraint2} specifies the energy budget constraint.
The left-hand side gives the sum of the energy consumed in the off state and active states.
The energy consumption of the $i^\mathrm{th}$ DP is given by the product of its power consumption $P_i$ and the active time allocated to it $t_i$.
Energy budget $E_b$ on the right-hand side is determined by energy allocation
techniques using the expected amount of harvested energy and battery
capacity~\cite{kansal2007power,bhat2017near}.
%
%
Finally, Equation~\ref{eq:constraint3} ensures that all active times
are non-negative.


\vspace{-1mm}
\subsection{Runtime Optimization Algorithm}

The solution of the proposed optimization problem
provides the duration of active times for each design points
such that they can collectively maximize the objective function in
Equation~\ref{eq:objective}.
It is important to solve this problem at runtime because the available energy budget is not known at design time.
Furthermore, the importance given to accuracy versus active time (i.e., $\alpha$) may change due to user preferences.

The optimization objective and the constraints in Equations~\ref{eq:objective}, \ref{eq:constraint1} and \ref{eq:constraint2}
are linear in the decision variables $t_i$ and $t_{off}$ for $1\leq i \leq N$.
Therefore, we use a procedure based on the simplex
algorithm~\cite{cormen2009introduction}, as
outlined in Algorithm~\ref{real_algorithm}.
It takes the energy budget $E_b$, Pareto-optimal DPs,
and the maximum number of iteration as input arguments.
The output is a vector with the
values of decision variables $t_i$ and $t_{off}$ for $1\leq i \leq N$ that
maximize the objective value.
We start the optimization process by constructing
a tableau with the initial conditions. The first row of the tableau describes
the objective function, while the other rows describe the constraints.
In each iteration of the procedure, we first find the pivot column by finding the column with the largest value in the last row of
the tableau.
Using the pivot column, we next find the pivot row in the tableau
in Line 8 of the algorithm.
Then, we update the tableau using the pivot column and row. The procedure is
terminated when all the entries in the last row are non-positive. In this case,
the pivot column is set as negative and the
optimal solution is returned.
\textit{Our implementation takes 1.5~ms on our IoT device
prototype with five design points}. We also observe that the algorithm takes 
only 8~ms for up to 100 design points. 
Since we run the optimization algorithm every hour, it takes
a negligible portion of the activity period and energy budget.

\vspace{-2.5mm}
\begin{algorithm}[h]
	\SetKwData{Left}{left}\SetKwData{This}{this}\SetKwData{Up}{up}
	\SetKwFunction{Union}{Union}\SetKwFunction{FindCompress}{FindCompress}
	\SetKwInOut{Input}{Input}\SetKwInOut{Output}{Output}
	\caption{The \textit{REAP} Procedure} \label{real_algorithm}
	\Input{Design points, energy budget $E_b$, max. iterations}
	\Output{Time allocated to each design point}
	\BlankLine
	\SetAlgoLined
	Initialize the $tableau$ with objective function and constraints  \\
	Add slack variables for inequality constraints \\
	\While {iter $\le$ max. iterations}  {
		$PivotCol \leftarrow$  $findPivotCol(tableau)$\\
		\If{$PivotCol < 0$}{\Return Optimal Solution}
		$PivotRow \leftarrow findPivotRow(tableau, PivotCol)$ \\
		Update the $tableau$ using the $PivotCol$ and $PivotRow$ \\
	}
\vspace{-1mm}
\end{algorithm}



\section{\hspace{-3mm} \large{Human Activity Recognition Case Study}}\label{sec:design_points}

\definecolor{Gray}{gray}{0.9}		
\newcolumntype{g}{>{\columncolor{Gray}}c}	
\renewcommand{\arraystretch}{0.8}
\begin{table*}[t]
	\footnotesize
	\centering
	\caption{\small{Accuracy, execution time, power and energy consumption of different human activity recognition application design points.}}
	\vspace{-4mm}
	\label{tab:design_points}
	\resizebox{\linewidth}{!}{
		\begin{tabular}{@{}ccg|cccc|cc|gg@{}}	
			\toprule
			\multicolumn{3}{c}{Design point description} \vline &
			\multicolumn{4}{c}{MCU exec. time distribution (ms)} \vline
			& \multicolumn{4}{c}{Per activity summary} \\ \midrule
			\begin{tabular}[c]{@{}c@{}}DP \\ no.\end{tabular} & Features &
			\begin{tabular}[c]{@{}c@{}}Accuracy\\ (\%)\end{tabular} &
			\begin{tabular}[c]{@{}c@{}}Accel. \\ features\end{tabular} &
			\begin{tabular}[c]{@{}c@{}}Stretch \\ features\end{tabular} &
			\begin{tabular}[c]{@{}c@{}}NN\\ classifier\end{tabular} & Total &
			\begin{tabular}[c]{@{}c@{}}MCU\\ energy\\ (mJ)\end{tabular} &
			\begin{tabular}[c]{@{}c@{}}Sensor \\ energy\\ (mJ)\end{tabular} &
			\begin{tabular}[c]{@{}c@{}}Energy\\ (mJ)\end{tabular} &
			\begin{tabular}[c]{@{}c@{}}Power\\ (mW)\end{tabular} \\ \midrule
			1 & \begin{tabular}[c]{@{}c@{}} Statistical acceleration,\\ 16-FFT
				stretch\end{tabular} & 94 & 0.83 & 3.83 & 1.05 & 5.71 & 2.38 &
			2.10
			& 4.48 & 2.76 \\ \midrule
			2 & \begin{tabular}[c]{@{}c@{}}Statistical y-axis accel.,\\ 16-FFT
				stretch\end{tabular} & 93 & 0.27 & 3.83 & 1.00 & 5.10  & 2.29 &
			1.43 & 3.72 & 2.30 \\ \midrule
			3 & \begin{tabular}[c]{@{}c@{}}Statistical x- and y-axis \\
				accel. (0.8~s), 16-FFT stretch
			\end{tabular} & 92 & 0.27 & 3.83 & 0.90 & 5.00 & 2.10 &
			0.84 & 2.94 & 1.82 \\ \midrule
			4 & \begin{tabular}[c]{@{}c@{}}Statistical y-axis\\
				accel. (0.6~s), 16-FFT stretch
			\end{tabular} & 90 & 0.14 & 3.83 & 1.00 & 4.97 & 2.09 &
			0.57
			& 2.66 & 1.64 \\ \midrule
			5 & 16-FFT stretch & 76 & 0.00 & 3.83 & 0.88 & 4.71 & 1.85 & 0.08 &
			1.93 & 1.20 \\ \bottomrule
			\vspace{-3mm}
		\end{tabular}
	}
\end{table*}

\textit{REAP} is broadly applicable to energy-harvesting IoT devices that operate under a fixed energy budget.
In order to illustrate the optimization results on a real example,
we employ human activity recognition, i.e., HAR, as a driver application.

\vspace{-2mm}
\subsection{\hspace{-1mm}Background and Baseline Implementation}\label{sec:har_background}

There is a steady increase in the use of wearable and mobile devices
for the  treatment of movement disorders and obesity-related
diseases~\cite{espay2016technology}.
This technology enables data collection while the patients perform their daily activities.
The first step in this effort is to understand what activity the user is
performing at a given time. 
For example, the gait quality of the patient cannot be checked unless we know the user is walking.
Therefore, HAR on mobile devices has recently attracted
significant attention~\cite{lara2013survey}.

\begin{figure}[t]
	\centering
	\includegraphics[width=1\linewidth]{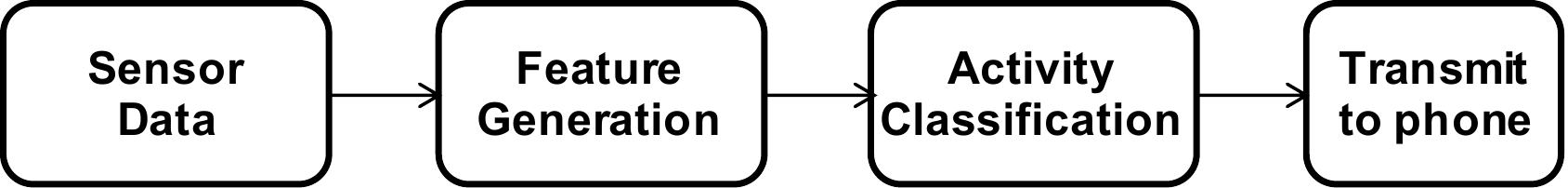}
	\vspace{-7mm}
	\caption{\small{Overview of the human activity recognition application.}}
	\label{fig:har_overview}
    \vspace{-4mm}
\end{figure}

We implement a HAR application on a custom prototype based on the
TI-Sensortag IoT board~\cite{TI_sensortag} and a passive stretch sensor.
Figure~\ref{fig:har_overview} shows an overview of our HAR application. It
starts with sampling of the accelerometer and stretch sensors. The streaming
sensor data
is fed to the TI-CC2650 MCU to generate the feature vector. The feature vector
is then processed by a parameterized neural network to infer the activity of
the user.
Finally, the inferred activity is transmitted to a host device, such as a
phone, using the Bluetooth Low Energy~(BLE) protocol.



\begin{figure}[h]
	\centering
	\vspace{-2.5mm}
	\includegraphics[width=0.98\linewidth]{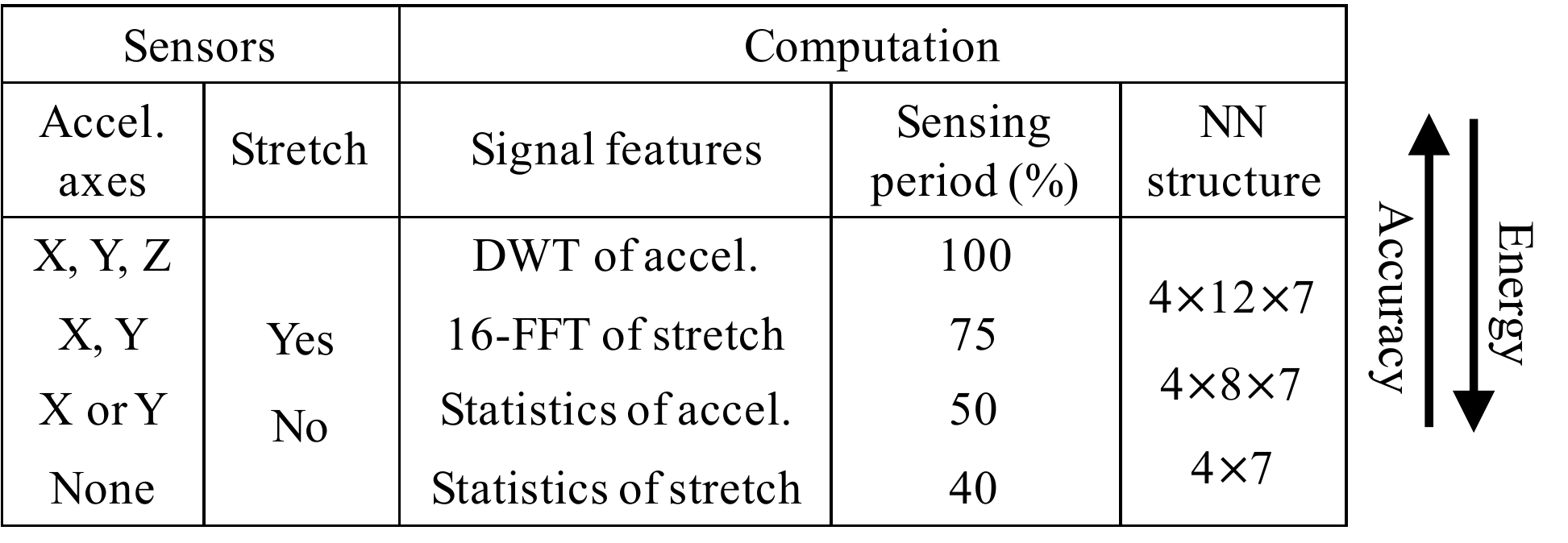}
	\vspace{-4mm}
	\caption{The knobs used to obtain design points with different energy-accuracy trade-offs.}
	\vspace{-3.5mm}
	\label{fig:dp_combinations}
\end{figure}

The energy consumption and recognition accuracy of HAR depends on the types of
sensors used, active time of the sensors, the type of features and the
complexity of the classifier.
Figure~\ref{fig:dp_combinations} shows the trade-off in energy and accuracy of
HAR for different
choices of sensors, features and classifiers. The left side of
Figure~\ref{fig:dp_combinations} shows the different configurations available for the sensors.
For instance, we can use all three axes of the accelerometer or turn off 
selected axes to lower the energy consumption.
In the extreme case, we can
turn off the accelerometer to completely eliminate its energy consumption.
Once we choose the configuration of the sensors,
we can choose the sensing period i.e. the time for which sensors are active for each activity duration.
By default, the sensors are turned on during the full activity duration.
Turning off the sensors early, such as after 50\% of the activity
duration, provides energy savings at the cost of missed data points, hence accuracy.
We can also control the complexity of the features to trade-off accuracy and energy consumption. Complex features, such as Fast Fourier Transform~(FFT) and
Discrete Wavelet Transform~(DWT), offer a higher accuracy at the expense of a higher energy consumption. In contrast, statistical features have a lower energy consumption, albeit with a lower accuracy.
Finally, the structure and depth of the NN classifier can be controlled to obtain further energy-accuracy trade-off, as illustrated in Figure~\ref{fig:dp_combinations}.


We exploit this trade-off between energy and accuracy to design 24 different DPs implemented on the TI-Sensortag based prototype,
as described in the following section.

\vspace{-1mm}
\subsection{Pareto-Optimal Design Points}
We design a total of 24 DPs by exploiting the energy-accuracy trade-off illustrated in Figure~\ref{fig:dp_combinations}.
We start by using all the axes of the accelerometer, generating complex
features, and using an NN classifier with 3 hidden layers, which provide the highest recognition accuracy.
Then, we progressively reduce the number of axes of the accelerometer and sensing period  to reduce the energy consumption of the DPs.
We always use the passive stretch sensor in our DPs, since it has a
low energy consumption.
There is a need for detailed accuracy and energy consumption characterization of each DP to obtain the Pareto-optimal design points. To find the accuracy of each
design point, we performed experiments with 14 different users.
We have obtained a total of 3553 activity windows from the experiments and labeled each window with the corresponding activity.
Each DP is designed using 60\% of this data for training, 20\% for validation and the remaining 20\% for testing.


\begin{figure}[t]
	\centering
	\includegraphics[width=0.80\linewidth]{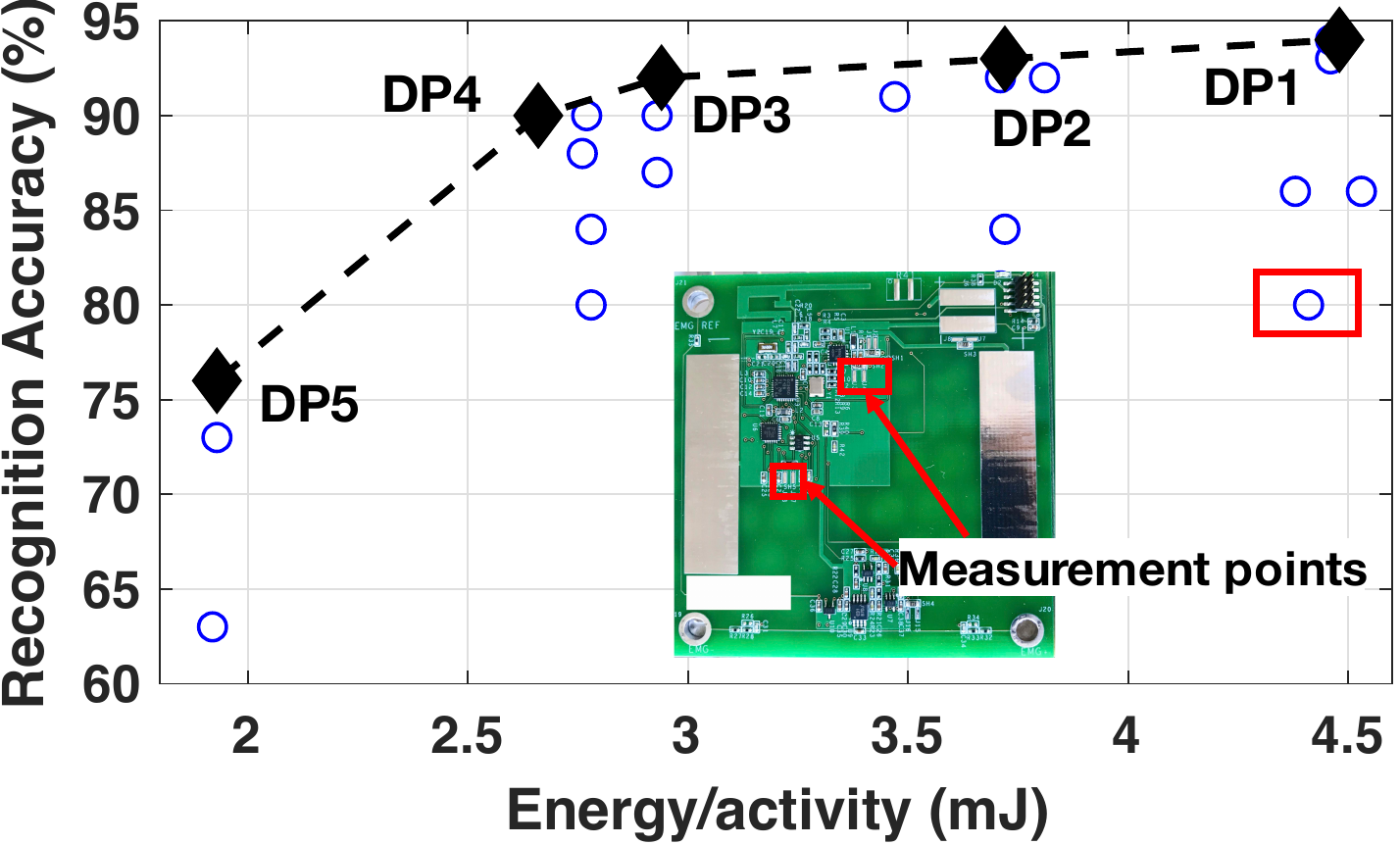}
	\vspace{-4mm}
	\caption{\small
		The energy-accuracy trade-off of
		various design points. The dashed line connects the selected five
		design points.}
	\label{fig:dp_plot}
	\vspace{-6mm}
\end{figure}

All 24 design points are implemented on our prototype
to profile the execution time and measure the power consumption using the test pads on our prototype.
%
Figure~\ref{fig:dp_plot} shows the recognition accuracy and energy per
activity for each design point. As expected, each DP offers a
unique energy-accuracy trade-off.
For example, DP1 shows the highest accuracy with the highest energy consumption while DP5 shows the lowest
recognition accuracy and energy consumption. However, some
design points do not offer any benefit in the energy-accuracy trade-off. For
example, the design point marked with a red rectangle is dominated by DP2, DP3
and
DP4. Hence, we consider five Pareto-optimal design points shown using black diamonds~(DP1 to DP5) for validating
the proposed REAP algorithm.
\textit{Table~\ref{tab:design_points} summarizes the details of the configuration, accuracy, execution time and energy for five Pareto-optimal DPs}.
Next, we provide a short description of the Pareto-optimal DPs.

\begin{figure}[t]
	\vspace{-1mm}
	\centering
	\includegraphics[width=0.68\linewidth]{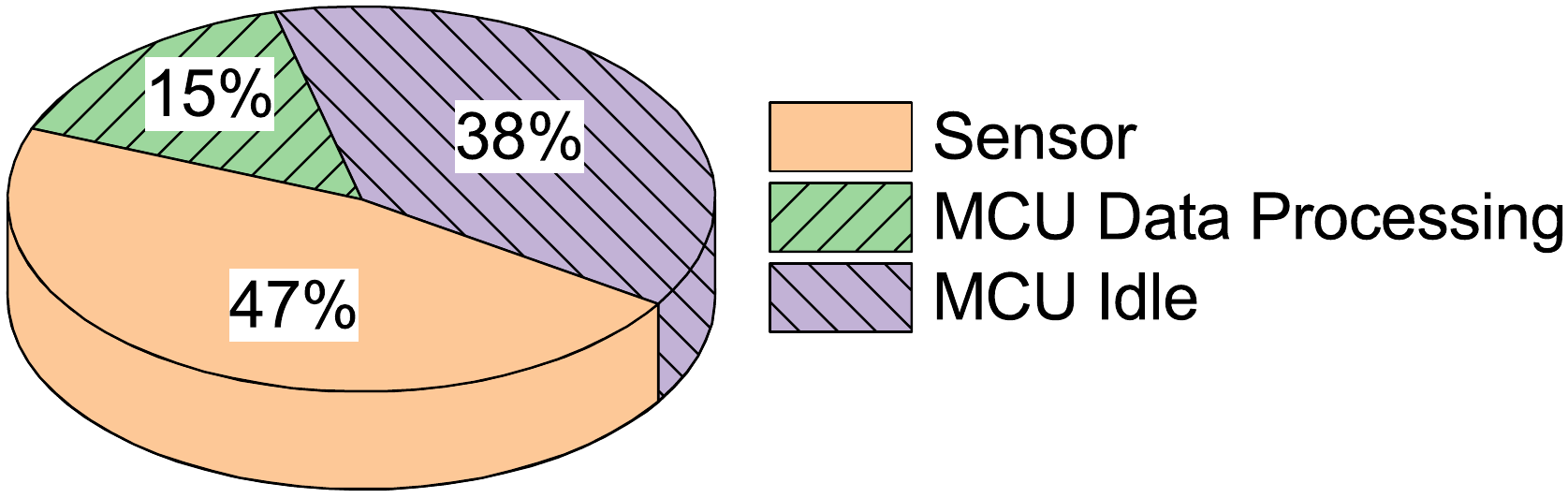}
	\vspace{-3mm}
	\caption{\small{\textcolor{black}{Energy consumption distribution of DP1
	over one-hour activity period $T_P$.
	Total energy consumption is 9.9~J.}}}
	\label{fig:dp1_energy}
	\vspace{-3mm}
\end{figure}

\vspace{1mm}
\noindent\textbf{Design Point-1~(DP1):}
DP1 offers the highest accuracy by utilizing all three axes of the
accelerometer for the entire activity window of 1.6~s.
It uses 16-FFT of the stretch sensor data and statistical features of the accelerometer, such as the mean and standard deviations.
DP1 leads to highest accuracy of 94\% at the cost of highest energy consumption 
of 4.48~mJ per activity.
The energy break-down in Figure~\ref{fig:dp1_energy} shows that about 47\% of the energy consumption is due to the sensors.
Thus, reducing the sensor activity is an effective mechanism to save energy.

\vspace{1mm}
\noindent\textbf{Design Point-2~(DP2):}
DP2 reduces the sensory energy  by utilizing only the y-axis of the
accelerometer along with the stretch sensor.
As depicted in Table~\ref{tab:design_points}, the energy consumption of the sensor is reduced to 1.43~mJ from 2.10~mJ.
It achieves an accuracy of 93\%, which is only 1\% lower than DP1.

\vspace{1mm}
\noindent\textbf{Design Point-3~(DP3):}
As shown in Figure~\ref{fig:dp_combinations}, reducing the sensing period leads to a lower energy consumption.
DP3 exploits this by sampling the x- and y- axes of the accelerometer for 50\% of each activity window, i.e., 0.8~s.
As a result, the energy consumption of the sensor is reduced to 0.84~mJ and the
total energy consumption of DP3 is reduced to 2.94~mJ per activity, while the
recognition accuracy drops to 92\%.

\vspace{1mm}
\noindent\textbf{Design Point-4~(DP4):}
DP4 is similar to DP3, except that sensing period of accelerometer is
further reduced to 40\% (0.6 s).
This reduces the energy consumption of DP4 to 2.66~mJ per activity with
recognition accuracy of 90\%.

\vspace{1mm}
\noindent\textbf{Design Point-5~(DP5):}
DP5 uses only the stretch sensor for data features to minimize energy consumption.
The energy consumption is reduced to 1.93~mJ per activity which is the lowest energy consumption among all our design points. However, it also shows the lowest recognition accuracy of 76\%.

\noindent\textbf{Offloading to a host:}
Finally, we note that the raw sensor data can be directly sent to a host device, such as a smartphone or server, for processing.
To assess the viability of this alternative, we implemented and measured its energy consumption.
Sending the raw sensor data over BLE consumes 5.5 mJ per activity without any 
significant increase in the recognition accuracy.
In contrast, transmitting just the recognized  activity consumes only about 0.38 mJ per activity.
Hence, offloading is not an energy-efficient choice.

\vspace{-1mm}
\section{Experimental Evaluation} \label{sec:experiments}

\vspace{-0.5mm}
\subsection{Experimental Setup} \label{sec:experimental_setup}
\vspace{-0mm}

\noindent\textbf{IoT device:} We use a custom prototype
based on the TI-Sensortag IoT platform~\cite{ TI_sensortag}
to implement the proposed design points.
The prototype consists of a TI CC2650 MCU, Invensense MPU-9250 motion sensor unit,
a stretch sensor and energy harvesting circuitry.
Sensors are sampled at 100~Hz and the MCU runs at 47~MHz frequency.
Power measurements from the prototype and data from 14 user subject studies
are used to obtain the 24 design points.
This data will be released to the public at \textit{blind-url}.

\noindent\textbf{Energy harvesting data:}
We use the solar radiation data measured by the NREL Solar
Radiation Research Laboratory
to obtain the energy harvesting profile from January 2015 to October
2018~\cite{NREL}.
We use the profile for each hour within this data to generate the energy budget.
These energy budgets are then used to evaluate \textit{REAP} and the static design points in Section~\ref{sec:case_study}.


\vspace{-1.5mm}
\subsection{\hspace{-2mm}Expected Accuracy and Active Time Analysis} 
\label{sec:impact_allocated_energy}
\vspace{-0.5mm}

We first analyze the results of the proposed optimization approach as a function of the allocated energy over one-hour activity period $T_P$. 
In the most energy-constrained scenario, 
the minimum energy required to run the energy harvesting and monitoring circuitry is 0.18~J. 
In the opposite extreme, 9.9~J energy is sufficient to run DP1, 
the most power hungry design point, throughout $T_P$. 
Therefore, we sweep the allocated energy starting with 0.18~J, 
and find the optimal active time of each DP using the proposed 
approach. 

\begin{figure}[t]
	\centering
	\includegraphics[width=0.95\linewidth]{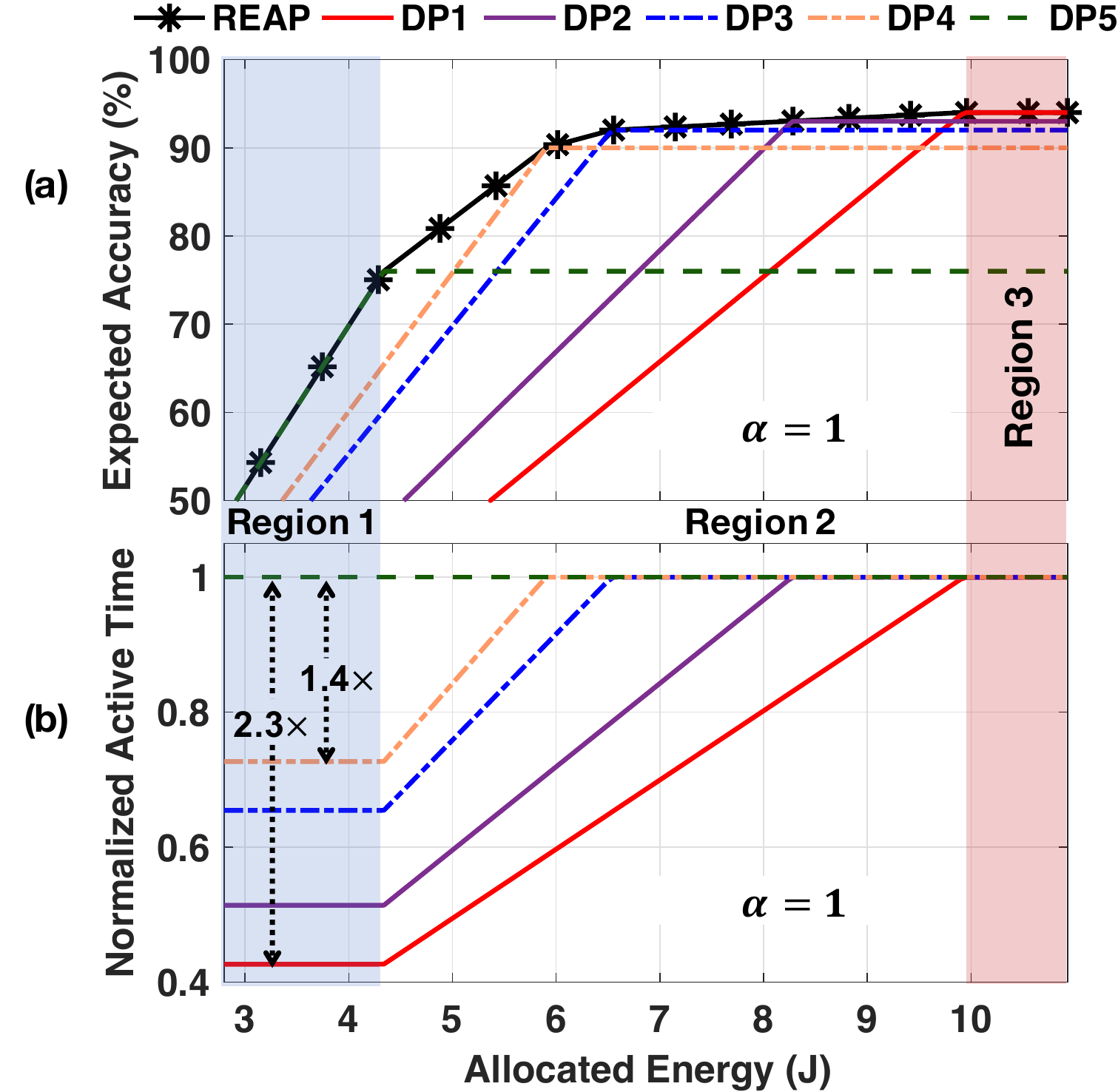}
	\vspace{-4mm}
	\caption{(a) Expected accuracy of \REAP and 
				design points.  
				(b) Active time of each DP normalized to \REAP.}	
	\vspace{-5mm}
	\label{fig:accuracy_act_time_Ec}
\end{figure}

Figure~\ref{fig:accuracy_act_time_Ec}(a) shows the expected accuracy ($\alpha = 1$) as function of the energy budget.  
The expected accuracy of all the design points approaches to zero 
when the energy budget is close to 0.18~J, 
since the device is almost always off. 
As the energy budget increases (Region 1), 
the accuracy of all DPs starts growing 
since they can become active. 
None of the design points can afford to stay 100\% active under the energy budget in Region 1. 
We observe that design points with lowest energy consumption (DP5) 
achieve significantly higher accuracy because they can stay in the 
active state much longer. 
\REAP successfully matches or exceeds 
the accuracy of DP5 under the most energy constrained scenario. 
When the energy budget goes over 4.3~J,
DP5 can remain active throughout the activity period but its recognition 
accuracy saturates. 
The other DPs benefit from more energy in Region 2, while \REAP outperforms all by utilizing them optimally. 
At 5~J energy budget, for example, \textit{REAP} utilizes 
DP4 42\% of the time and DP5 for 58\% of the time to optimize the expected 
accuracy.
Finally, all design points can remain active throughout the activity period 
when the energy budget is larger than 9.9~J.
Hence, their accuracy saturates, and \textit{REAP} reduces to DP1 beyond this 
point.
In summary, \REAP consistently outperforms or matches 
the accuracy of all individual DPs
by utilizing \textit{multiple DPs} optimally. 

%
The active time of each DP \textit{normalized to} \REAP is plotted in Figure~\ref{fig:accuracy_act_time_Ec}(b). 
DP5 is expected to have the longest active time since it has the least energy consumption. 
\textit{REAP} successfully matches its active time in all the regions.
In Region 1, \REAP also achieves 2.3$\times$ larger active time 
compared to DP1 while providing significantly better accuracy. 
\textit{REAP} consistently provides longer active times compared to DP1, DP2 and DP3
until the energy budget becomes large enough 
to sustain them throughout the activity period $T_P$.
%
%
%
%

%

\vspace{-1mm}
\subsection{Accuracy -- Active Time Trade-off Analysis}


\textcolor{black}{
Next, we analyze how \REAP can exploit trade-off between the accuracy and active time using the parameter $\alpha$ in objective function $J(t)$ in Equation~\ref{eq:objective}. 
Since Section~\ref{sec:impact_allocated_energy} considered the expected accuracy ($\alpha$= 1), 
this section considers $\alpha > 1$, which gives more emphasis for higher accuracy.}

\begin{figure}[t]
	\centering
	\includegraphics[width=0.89\columnwidth]{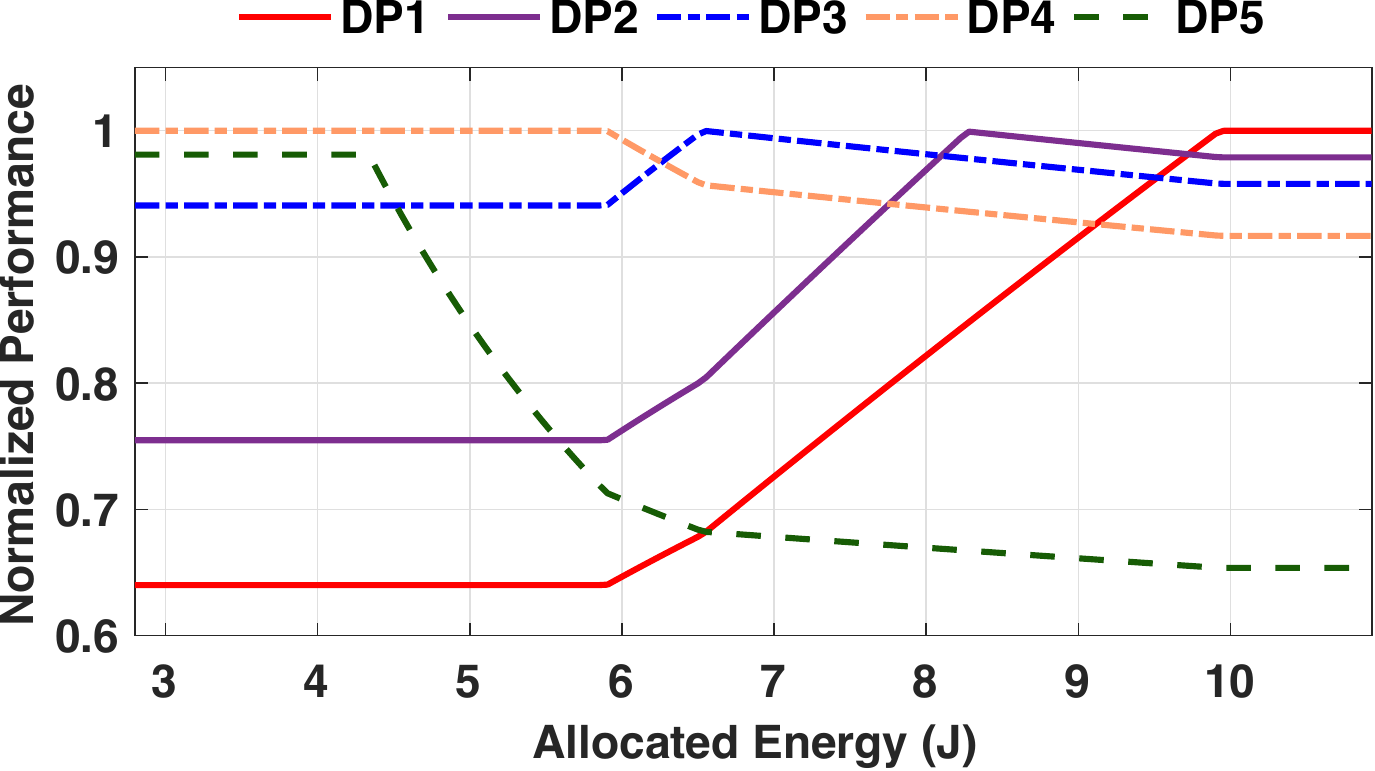}
	\vspace{-4mm}
	\caption{The objective value $J(t)$ given in 
				Equation~\ref{eq:objective} of static design points is normalized over 
				$J(t)$ of \textit{REAP} with $\alpha$= 2.}
	\label{fig:alpha2_4}
	\vspace{-4mm}
\end{figure}

As a representative example, Figure~\ref{fig:alpha2_4} shows the comparison of objective values of the five design points with \REAP when $\alpha$ is set to 2.
\REAP always achieves higher performance than the lowest energy design DP5, since accuracy is given higher weight. 
The difference between \REAP and  DP5 increases further as $alpha$ grows. 
When the energy budget is less than 6~J, DP4 outperforms all the other DPs, 
while \REAP successfully matches it. 
In contrast, DP1, DP2, and DP3 have a very low performance, since they are mostly in the off state. 
When the energy budget exceeds 6~J, there is sufficient energy to provide a higher accuracy, but DP4 cannot exploit it.
Hence, the higher accuracy design points become affordable and start outperforming DP4 one by one. 
Notably, \REAP consistently outperforms or matches the static DPs, as we have 
also observed in Figure~\ref{fig:accuracy_act_time_Ec}. 
For example, DP3 is able to provide the same performance as \REAP when the energy budget is 6.5~J. 
As the energy allocation increases beyond 6.5~J, 
\REAP starts outperforming DP3 by optimally switching between DP1, DP2 and DP3. 
This trend continues until the energy allocation reaches 9.9~J, 
beyond which there is sufficient energy to support DP1 alone. Thus, \REAP reduces to DP1 in this region. 
In summary, \REAP exceeds or matches the 
performance of any individual DP.

\vspace{-1.5mm}
\subsection{Case Study using Real Solar Energy Data}
\label{sec:case_study}
\vspace{-0.5mm}

\begin{figure}[t]
	\vspace{-3mm}
	\centering
	\includegraphics[width=1\linewidth]{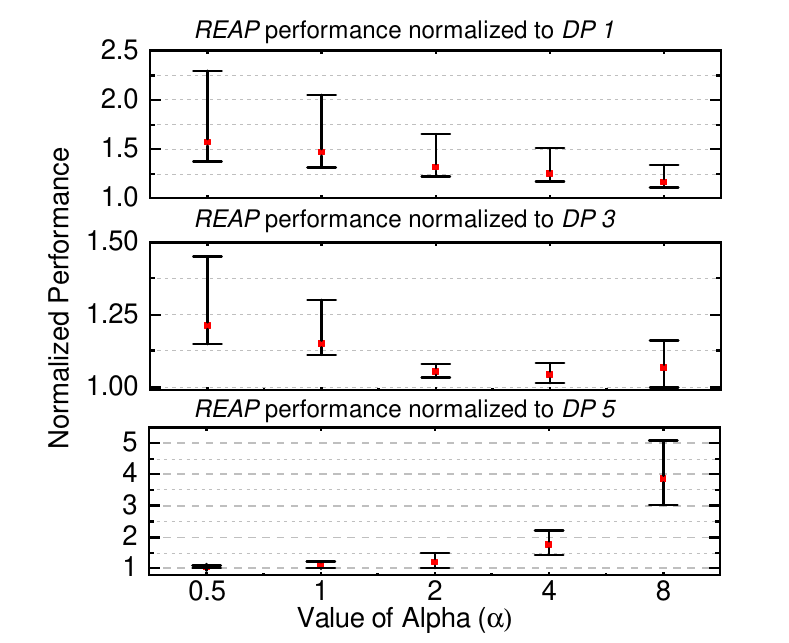}
	\vspace{-5mm}
	\caption{Performance (i.e., J(t) ) achieved by \REAP normalized to DP1, 
	DP3, and DP5 during the month of September 2015. Error bars represent the 
	range of improvement.}
	\label{fig:one_month_plot}
\end{figure}

In this section, we evaluate \REAP under the real solar radiation data measured by NREL Solar Radiation Research Laboratory 
at Golden, Colorado. 
This data is used to calculate the amount of energy that can be harvested by a flexible solar cell~\cite{sp3-37} on our prototype.
Using the harvested energy budget, we compare the performance of \REAP 
against the static DPs over an entire month. 
Figure~\ref{fig:one_month_plot} shows 
the performance of \REAP normalized to DP1, DP3, and DP5 as a function of $\alpha$. 
Due to space limitation, we plot the DPs with the highest performance (DP1), lowest energy (DP5), 
and best trade-off (DP3). 
Our gains with respect to DP2 and DP4 are larger than that of DP3.

When active time is emphasized in the objection function ($\alpha=0.5$), 
\REAP outperforms DP1 by 1.4$\times$--2.2$\times$ with an average improvement of 1.6$\times$ across the month. 
DP1 suffers the most in this case as it has the largest energy consumption among all the DPs. 
Since accuracy becomes more important with larger $\alpha$,  
the improvement of \REAP over DP1 reduces. 
However, we still obtain 1.1$\times$--1.3$\times$ improvement even for $\alpha=8$. 
We observe a similar trend in improvements for \REAP for DP3 as well. 
The improvement is 1.1$\times$--1.4$\times$ for $\alpha=0.5$, and it 
gradually decreases with larger $\alpha$. 
The improvements over DP3 are relatively lower, since DP3 offers the best trade-off between 
energy consumption and accuracy among our Pareto-optimal design points.

Finally, we see that the improvements over DP5 follow the opposite trend. 
When $\alpha = 0.5$, DP5 is able to match the active time of \REAP due to its lower 
energy consumption.
However, the performance of DP5 diminishes severely with increasing $\alpha$. 
In summary, \REAP can 
provide a higher performance than any individual design point under any optimization objective. 
If the user needs a higher accuracy, \REAP can 
successfully adapt to new requirements.  
This can be utilized by the IoT device to tune 
its performance as user needs change.

\vspace{-1mm}
\section{Conclusions} \label{sec:conclusions}

This paper presented a runtime accuracy-active time optimization technique for 
energy-constrained IoT devices. 
The proposed approach dynamically chooses design points with different 
energy-accuracy trade-offs to co-optimize the accuracy and active time under 
energy budget constraints. 
To demonstrate the effectiveness in a realistic setting, we implemented a human 
activity recognition application on a custom IoT prototype. 
We presented five Pareto-optimal design points with different energy-accuracy trade-offs.
We achieve 46\% higher expected accuracy and 66\% longer 
active time compared to the highest performance design point, 
and 22\% to 29\% higher accuracy than low-power design points without sacrificing the active time.

\vspace{-1mm}


\footnotesize{\bibliographystyle{abbrv}}



\end{document}